\documentclass[usenatbib]{mn2e}
\usepackage{graphicx}
\usepackage{array}
\usepackage{multirow}
\usepackage[font=small,format=plain,labelfont=bf,up,textfont=normal,up]{caption}
\usepackage{float}
\usepackage{txfonts}

\title[Spectro-photometric variability of quasars caused by
lensing of diffuse  massive substructure]{Spectro-photometric variability of quasars caused by
lensing of diffuse  massive substructure:\\
Consequences on flux anomaly and precise astrometric measurements}

\author[L. \v C. Popovi\'c \& S. Simi\'c]{L. \v C. Popovi\'c$^{1,2,3}$\thanks{E-mail:
lpopovic@aob.rs},S. Simi\'c$^{2,4}$\\
$^{1}$Astronomical Observatory, Volgina 7, Belgrade, 11160, Serbia\\
$^{2}$ Isaac Newton Institute of Chile, Yugoslavia Branch\\
$^{3}$ Faculty of Mathematics, University of Belgrade\\
$^{4}$ Faculty of Science, University of Kragujevac, Radoja Domanovi\'ca 12, 34000 Kragujevac, Serbia\\
}

\begin{document}

\date{Accepted 18.03.2013. Received February 2013}


\maketitle

\label{firstpage}

\begin{abstract}
We investigate the spectro-photometric variability of quasars due to lensing of small mass substructure (from several tens
to several hundreds solar masses). The aim of this paper is to explore the milli/microlensing influence on
the flux anomaly observed between images of a lensed quasar in different spectral bands and possible influence of small mass
structure lensing of non-macrolensed quasars. We find that spectro-photometric variability may be also caused by
lensing of small mass diffuse structure and can produce the flux anomaly which is sometimes seen in
different images of a lensed quasar. Additionally, we found that the lensing by small mass diffuse structure
may produce significant changes in photo-center position of a quasar, and sometimes can split or deviate images of one
source that can be detected as separate from the scale from 0.1 to several milliarcseconds. This can be
measured ith Gaia-like space astrometric missions. We point out a special case where a low redshifted deflector ($z_d\sim0.01$) is
lensing a high redshifted source, for which the variability in the flux and photo-center (several milliarcseconds)
may be detected on a relatively short time scale.
\end{abstract}


\section{Introduction}
\label{sec:int}

The light from a distant quasar (QSO), along its path toward the Earth, can be perturbed by compact objects, as e.g. galaxies,
stars in galaxies, stellar clusters, intermediate mass compact objects (IMCOs with $10^{2-4} M_\odot$) and  cold dark matter
(CDM) structure. This can cause magnification in the luminosity of a QSO and in its photometric position, the so called lensing effects. The gravitational lensing can be strong or weak, depending on the position of a quasar and object (in the line of sight of an observer) which acts as a lens, if the line of sight from the observer to the source lie very close to the center of a massive object (galaxy cluster, galaxy, CDM, stellar cusp, etc.) then we expect a strong lensing effect. The strong lensing of QSOs can cause a significant wavelength dependent amplification and affect the photocentric position spectrum \citep[see e.g.][]{og05,co05,tr10,Erick11}. Depending on the image angular separation, the strong lensing can be divided into several categories from which
three are the most used \citep[see][]{za10,tr10}: macrolensing ($>0.1$ arcsec), millilensing ($10^{-3}$ arcsec) and microlensing ($10^{-6}$ arcsec). Usually, millilensing and microlensing are considered to be present in the images of a macrolensed QSO, however the effect of micro/millilensing may be
present in QSOs even if they are not macrolensed, i.e. the line of sight from the observer to source does not lie very close to the center of a massive galaxy, but still a compact massive object from the galaxy and stars can affect the QSO light \citep[see e.g.][]{zah04,za07}. There is a number of QSOs observed through a galaxy stellar disc \citep[see e.g.][]{me10}. The possibility that QSOs are micro/millilensed seems to be higher than macrolensed. As e.g. \cite{zah04} found that the optical depth for gravitational micro/millilensing caused by cosmologically distributed deflectors could be significant and could reach from 0.01 to 0.1 for QSOs with  $z>2$. Therefore, micro/millilensing effect may be often present not only in the case of lensed QSOs, but in a number of non-lensed objects. It is well known that micro/millilensing can be used for investigation of the structure of QSOs as well as lensing objects \citep[see e.g.][for review]{tr10,za10,sc10}, and especially they present a tool for detection of dark matter subhalos \citep[see, e.g.][etc.]{in05,ch07,za07,za10,Erick11}.

Mainly two effects of micro/millilensing are used to investigate structure of quasars and lensed objects: (i) a magnification that may be wavelength dependent due to complex structure of QSOs \citep[see e.g.][]{pop05,jov08,sl12,st12}, and in a such way producing a continuum (line) flux anomaly in the images of a macrolensed QSO \citep[see e.g.][etc.]{bl05,po06,kr11,po12}; and (ii) a slight wobbling of the observed source position caused by microlensing effect, called astrometric microlensing (photocenter variations), that presents a photo centroid displacement of the source image during the microlensing event in the lens plane \citep[see e.g.][etc.]{Hog95,Walker95,Miyamoto95,Dominik00,Lee10}. In the above papers it has been shown that for a single lens and a single source configuration, centroid motion during the microlensing event follows, more or less, an elliptical dependence on the impact parameter of the source and lens.

Astrometric microlensing has been explored by a number of authors covering different regimes. Most of them discussed cases where lenses are stars in the Milky Way or dark  matter objects in Galactic halo, acting on background stars either in the Magellanic Clouds or in nearby galaxies, as e.g. M31 \citep[see e.g.][etc]{Miralda96,Mao98,Boden98,Goldberg98,Paczinski98,Han99,Safizadeh99,Dominik00,Delplancke01,Belokurov02,Dalal03,Lee10,pr11,ya12}. The effects of Milky Way stars on background quasars were studied in several papers \citep[see e.g.][]{Hosokawa97, Sazhin98, Homna02, Sazhin11} and as well as the effect of local dark matter subhalos \citep[see e.g.][]{Erick11}. There are also studies that investigate astronomical lensing in intervening galaxy \citep[see][]{Williams95,ch07}. However, changing in magnification at different wavelengths (flux anomaly due to micro/millilensing) in combination with astrometric microlensing is rarely investigated.

A new class of satellites will be able to conduct precise measurements of an object photo-center in different spectral ranges. For instance, the satellite Gaia is equipped with most sophisticated instruments with astonishing capabilities, producing the accuracy better then $\rm 20\mu as$ in wide energy range $\rm 330 \ nm-1050\ nm$. The primary objective of this mission is to create a three-dimensional model of our Galaxy, but also to be used in other branches of astronomy, such as detection of extra-solar planets, brown dwarfs, asteroids, exploding stars, testing the Einstein theory, etc. One of the main tasks of this mission is the definition of fixed celestial referential frame, the Gaia realization of the International Celestial Reference System (ICRS; \cite{Mignard02}). To reach this purpose, a large number of QSOs ($5000-10 000$) uniformly distributed over the sky is required. QSOs are Active Galactic Nuclei (AGNs) located at cosmological distances. This makes them to a perfect candidate for this purpose as they are objects without or with negligible proper motion. The QSO sample dedicated to this task should exhibit absolutely no proper motion and be completely free from any contaminant (non QSO objects), while tiny apparent astrometric motions due to inner structure flux variations \citep[][]{pop12} or (micro)lensing are always possible \citep[][]{Treyer04}. As we noted above, micro/millilensing may be present also in not macrolensed QSOs. In the paper \cite{pop12} we investigate photocentric variability of quasars caused by variations in their inner structure, and here we give our investigation of the influence of the micro/millilensing of diffuse small mass structure (like an open stellar cluster) on photocentric variation of QSOs that can also have consequence for Gaia measurements.

The aim of this paper is to investigate spectro-astrometric variability of quasars caused by micro/millilensing taking into account the complex emitting structure of a QSO. Additionally we consider the lens as a bulk of stars concentrated on a relatively small surface, as e.g. star clusters, simulating a massive diffuse lensing structure that contains between several tens to several hundreds of solar mass stars. We explore astrometric microlensing in combination with spectral variation from the UV to IR spectra band. This could be of particular interest since both depends on specific the properties of a lens and a source \citep[see][]{ke09}.

The paper is organized as following: In the next section we present the source and lens model; in \S 3 we give results of our simulations and discuss results, and finally in \S 4 we outline our conclusions.

\section{The source and lens model}

To investigate photocentric variation in combination with possible wavelength dependent flux amplification, one has to take into account the complex structure of QSOs, but also consider the lens structure. Here we assumed that the radiation in the UV and optical band is mainly coming from an accretion disc. On the other side, one cannot expect that a single star in galaxies which are redshifted more than z$\ge$0.01 can produce significant astronomical microlensing, therefore we considered a massive diffuse structure containing a group of stars distributed in a small surface.

\subsection{Source model -- stratified emission disc in QSOs}
\label{sec:srcmodel}

An AGN has a complex inner structure, meaning that different parts have emission in different spectral bands, i.e. a wavelength dependent dimensions of different emission regions is present in AGNs \citep[see e.g.][]{pop12}. It is widely accepted that the majority of radiation is coming from the accelerated material spiraling down toward the black hole in a form of an accretion disc. The radiation from the disc is mostly thermalized from outer regions $R_{out}$ to the center, with slight exception at inner radius, i.e. very close to the black hole, where Compton upscatering due to the high mass accretion could have a significant role \citep[see][]{Done11}.

To take into account the spectral stratification of the source we accepted a model of a standard relativistic, optically thick, geometrically thin, black body disc model \citep[see, e.g.][]{Pringle72,Shakura73,Novikov73} using the effective temperature as a function of the radius \citep[][]{Krolik98}:

$$T\propto R^{-3/4}(1-(R_{in}/R)^{1/2})^{1/4},$$

\noindent where $R_{in}$ is the inner disc radius. At larger radius this equation could be reduced to $T\propto R^{-\beta}$, where in the standard model $\beta = 3/4$. Here we are going to investigate the microlensed radiation from the UV to IR (not in X-ray), therefore we adopted the thermal emission as the main mechanism.

\begin{figure}
\includegraphics[width=8.5cm]{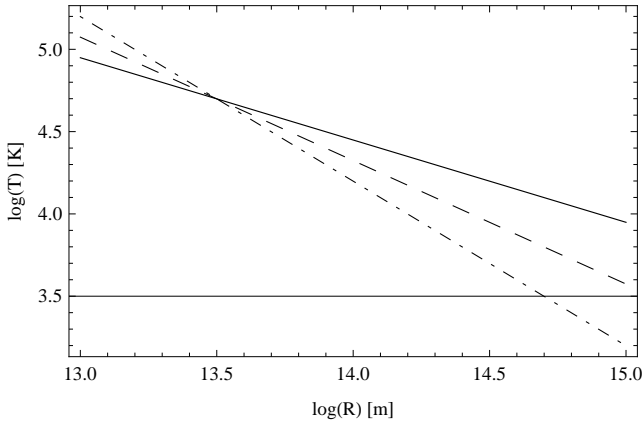}
\caption{Temperature as a function of the position of emitting material in the disc (in logarithmic scale) for different $\beta$
parameters: $\beta=0.5$ -- full line, $\beta=0.75$ -- dashed line, and $\beta=1.0$ -- dash-dotted line.}
\label{fig_TodR}
\end{figure}

Taking into account the temperature gradient along the disc we adopted the formalism similar as in \cite{Poindexter08}, where the luminosity of a small surface element at the arbitrary position in the disc is proportional to the surface energy density and magnitude of emitting surface \citep[][]{Poindexter08}

\begin{equation}
\label{eqn:source}
dL(\lambda, R)\propto \frac{dS}{\lambda^{5}} (exp(\frac{hc}{\lambda k \alpha(\beta) R^{-\beta}}-1))^{-1}.
\end{equation}

\noindent where $dS$ is the surface element of the source. We replace $T$ in the expression for the energy density with the distance $R$ as shown in Fig. \ref{fig_TodR} and compute the proportionality coefficient as

\begin{equation}
\label{eqn:T0R0}
\alpha=T_0 R_0^{\beta},
\end{equation}

\noindent where $T_0$ is the temperature at the distance $R_0$. To compute the spectral energy distribution (SED) for the disc configuration we integrate over the whole disc area:

\begin{equation}
\label{eqn:source_int}
L(\lambda)\propto \int_{S_{disc}}^{}dL(\lambda,R)
\end{equation}

\begin{figure}
\includegraphics[width=8.5cm]{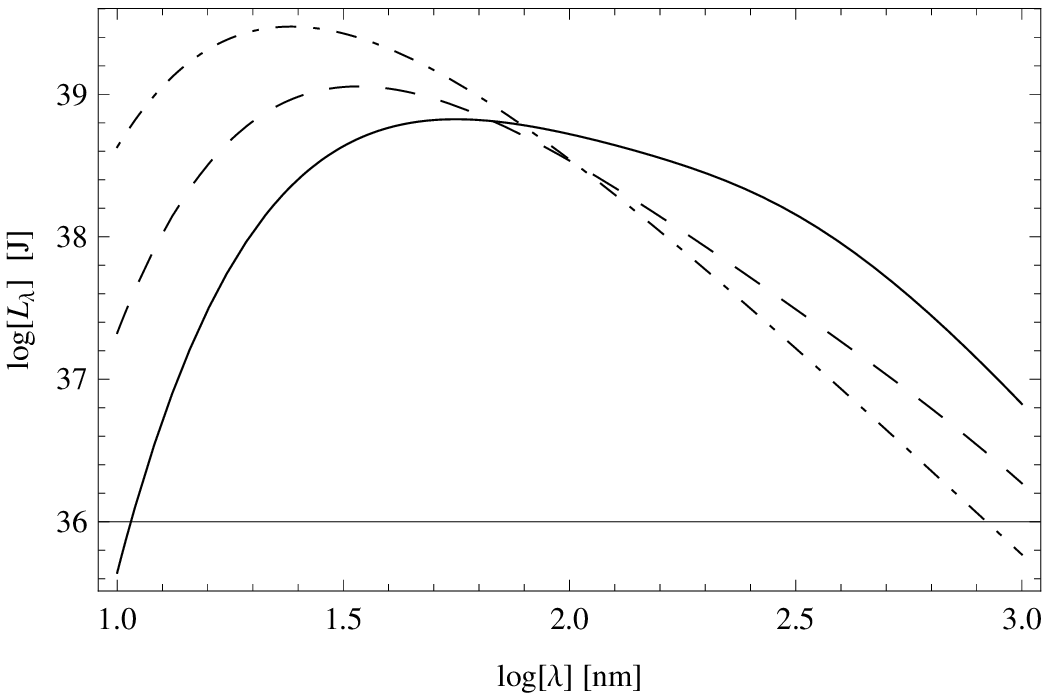}
\includegraphics[width=8.5cm]{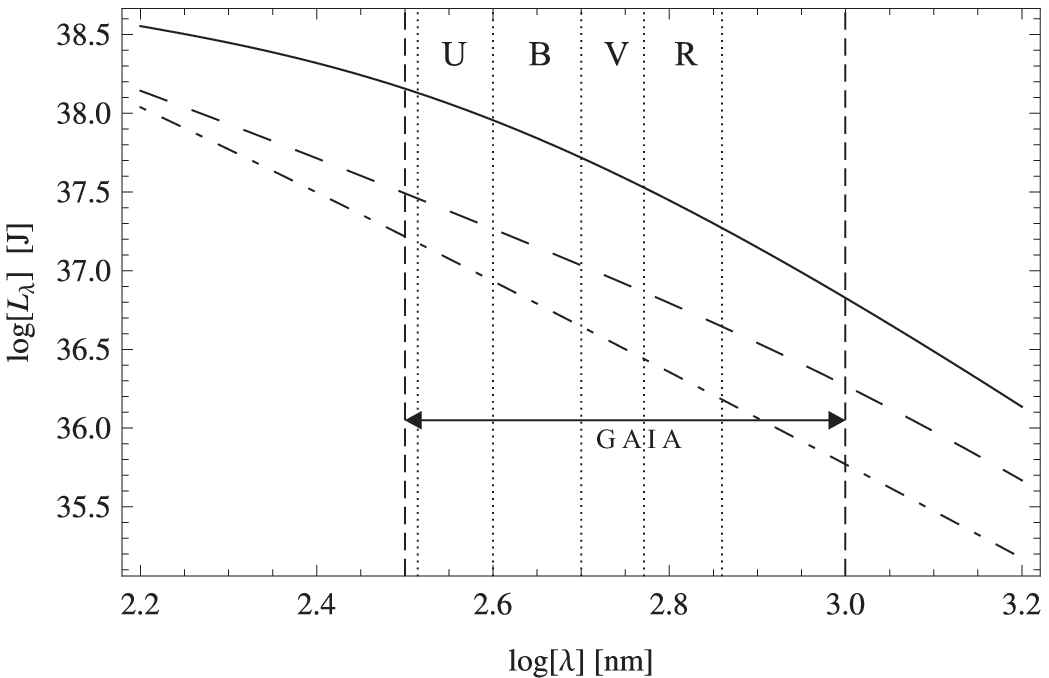}
\caption{The SED of a modeled source for three values of $\beta$ (notation is the same as in Fig. 1). Upper: For a wide spectral range. Bottom: For U, B, V, R filters (separated by dotted verticals) which are considered in the paper and also the spectral range which will be covered by Gaia is also shown.}
\label{fig_source}
\end{figure}

Using Eq. \ref{eqn:source_int} we calculated the SED of the source (see Fig. \ref{fig_source}). As it can be seen in Fig. \ref{fig_source} (up), the  SED  depends on value of the parameter $\beta$, and for  $\beta=0.5$ (full line) the SED has a maximum at lower energy than other two SEDs (with $\beta=0.75$ and 1, dashed and dash-dotted line, respectively). An increase of the source emission is noticeable at higher energies for higher values of the parameter $\beta$.

The inclination of the disc with respect to the observer could be incorporated with additional $cos(i)$ (\emph{i}-inclination angle). In all cases we considered a face-on disc ($i=0$).

Using the disc model for the continuum emission of QSOs we were able to model emission in the spectral range that will be covered by Gaia (see Fig. \ref{fig_source}, down). Moreover, such model allowed us to explore, both flux anomaly and photocenter variation in different spectral bands during a micro/millilensing event.

\subsection{Lens model -- diffuse massive substructure}
\label{sec:mlmodel}

As we noted above, the light from distant QSOs may cross far enough from a galaxy that is not macrolensed, but still may be micro/millilensed by structures which belong to the galaxy. Single star, or randomly distributed stars in the galaxy will probably act as a weak lensing, that is out of the scope of this paper.

However, the light may cross near the center of some massive structures which have significant surface density that can produce strong (micro/milli) lensing effect. As widely accepted, galaxies can contain substructures with few tens to few hundreds stars which are usually addressed as the open star clusters. They can be irregular in shape with dimension of a few parsecs or less.
Here we will discuss effects of group of stars usually referred as an open star cluster that contains from a few tens to few hundreds Sun like stars.

Such substructures contain enough stars to be able to produce the gravitational lensing on milli/micro scale. Here we investigate such structure as a possible lens. For magnification map simulation we used the ray-shooting technique, and first we describe the technique in a few sentences.

\subsubsection{Generation of micro/millilensing magnification maps}
\label{sec:gen_mag_map}

\begin{figure}
\includegraphics[width=9cm]{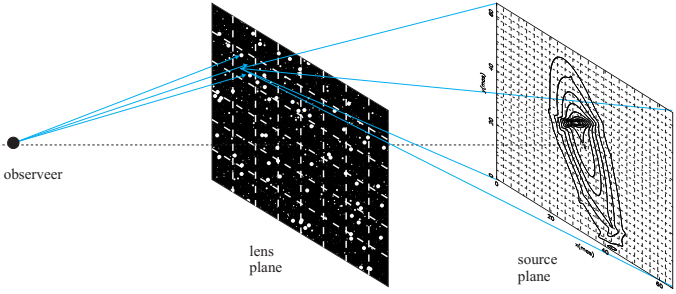}
\caption{A schematic representation of the ray-shooting technique in which a beam of light rays is shot from the observer
through the lens plane towards the source. In the source plane we sketch the source with concentric curved lines.}
\label{fig_ray}
\end{figure}

A distribution of stars in the lens plane generates microlensing magnification map in source plane which could be computed by the ray-shooting technique \citep[][]{Kayser86,Schneider86,Schneider87,Treyer04}. This technique proposes to follow a ray of light from the observer point passing towards the lens and falling on the source plane.
Light rays initiated with different angles close to the observer, will have different deflection angles at the lens plane and reach the source plane at different points. We place the grid in the source plane and count the number of light rays falling onto particular grid cells. In that way, the unique map is created representing the current star distribution in the lens plane. This map is called the magnification map and can be used to determine the amount of magnification of source imposed by the lens. In Fig. \ref{fig_ray} we present a sketch of the ray-shooting technique.

\subsubsection{Distribution of deflectors in the lens plane}

A microlens is often defined as a random distribution of sun-like stars with masses equal to $M_{\odot}$, over the lens surface.
This distribution is determined by the parameter $\kappa$, which presents average surface density. In nature, stars are mostly
distributed in clusters with different shapes. Therefore, considering the discussion at the beginning of \S \ref{sec:mlmodel}
here we used a random star distribution with the application of a scale parameter $\chi$, which could allow us to
change the compactness of the star cluster. Also, we can simulate density perturbation in a form of star groups in order
to produce galaxy substructures mentioned in chapter \S \ref{sec:int}. Additionally, we have adopted a thin lens
approximation, which assumes that all stars belonging to the lens are placed in one plane. This approximation is acceptable
when the lens and source are at cosmological distances, so we neglect the dimensions of the lens itself.

First, we randomly distributed a number of solar mass stars $N_{lens}$ in the circular area with angular diameter $d_l$. In the source plane of a squared shape with dimension $d_s$, we considered a grid of $N \times N$ pixels, where one pixel has dimension of $d_s/N$. The deflection angle has been calculated using the gravitational potential in each point of the lens plane as:

\begin{equation}
\label{eqn:alpha}
\vec{\alpha}=\nabla\psi
\end{equation}

For the calculation of the gravitational potential we solved the Poisson equation $\nabla^2\psi=2\kappa$ in the lens plane, and we used the lens equation for computing the coordinates of points (elements of the source plane grid) where light ray will fall. This process has been repeated a number of times for different incoming angles in order to produce the light rays distribution on the source plane, i.e. to generate the magnification map. The magnification of a point on the map is higher if more light rays fall on it. Those light rays are then distributed over all pixels in the source plane, and the number of rays per pixel is proportional to the magnification due to microlensing at that pixel in the source plane.

Using a magnification map created in described way in the source plane, we considered a source of light of a finite dimension $d_{src}$, that is also divided in a number of pixels. We accounted only for the light rays falling on the source itself or better to say rays emitted from the source surface. That radiation is additionally increased by the amount defined with distribution of caustics of magnification map in particular pixels. The emission of source pixels is determined by it's SED. Depending on the lens mass distribution, in the scope of the ray shooting technique, there can be more then one pixel in the lens plane where light rays sent from the observer penetrate and fall on the particular pixel in the source plane. Finally, any pixel of source has a collection of pixels in the lens plane that depends on it's SED. This procedure allowed us to create an image of the source in the lens plane in considered spectral range. Luminosity of the lens plane pixel $L_{pix}^{lens}(\lambda)$ is higher if there are more light rays passing through it and mathematically presents a sum of SED's for all passed light rays:

\begin{equation}
\label{eqn:pix_sed}
L_{pix}^{lens}(\lambda)=\sum_{nrays}L(\lambda,R).
\end{equation}

We calculated the centroid shift of the image for a spectral filter as:

\begin{equation}
\label{eqn:cent_shift}
D_{cs}(F)=\frac{\int_{F}\sum_{npix}x_{pix}L_{pix}^{lens}(\lambda) d\lambda}{\int_{A}\sum_{npix}L_{pix}^{lens}(\lambda)d\lambda}
\end{equation}

\noindent where $F$ denotes integration for a particular (U, B, V, R) spectral band and A is the whole energy range.

The magnification for a particular source image is computed as the ratio of the  luminosity for all pixels in a spectral
range with the luminosity in the same spectral range without lens influence, as:

\begin{equation}
\label{eqn:UBVR_magnification}
m(F) = \frac{\int_{F}\sum_{npix}L_{pix}^{lens}(F)d\lambda}{\int_{F}\sum_{npix}L_{pix}^{nolens}(F)d\lambda}.
\end{equation}

The relevant length scales for microlensing is the dimension of the Einstein Ring Radius (ERR) in the lens plane, defined as:
\begin{equation}
\label{eqn:xi0}
\xi_0=\sqrt{\frac{4Gm}{c^2}\frac{D_d D_{ds}}{D_s}},
\end{equation}
and its projection in the source plane is:
\begin{equation}
\label{eqn:err}
ERR=\frac{D_s}{D_d}\xi_0=\sqrt{\frac{4Gm}{c^2}\frac{D_sD_{ds}}{D_d}}
\end{equation}
\noindent where $G$ is the gravitational constant, $c$ is the speed of light, $m$ is the microlens mass. We adopted standard notation for cosmological distances to the lens $D_d$, source $D_s$ and between them $D_{ds}$.

Time scales for microlensing event is described in detail in the book on gravitational microlensing \citep[see][]{sch92,Zah97,Pett01}. Based on the sizes of the source ($R_{src}$) and caustic ($r_{caustic}$) pattern we distinguish two cases, when $R_{src}>r_{caustic}$ and $R_{src}<r_{caustic}$ (presented in detail in \cite{jov08}). Both cases could be expressed in a single form:

\begin{equation}
\label{eqn:time_scale}
t_{crossing}=(1+z_d)\frac{R}{v_{\perp}(D_s/D_d)}
\end{equation}
\noindent where $R$ replaces the $R_{src}$ or $r_{caustic}$. Here we used a simple approach, since we considered that a microlensing event duration corresponds to the time needed for crossing over the caustic network created by the lens. In this way the dimension of the caustic patterns in the magnification map determine the total time scale for particular event, and it can be computed by using the Eq. \ref{eqn:time_scale}, with the $R$ replaced by the dimension of map $r_m$. We used already introduced comoving distances $D_s$ and given map dimensions in ERRs to calculate maps linear dimensions, and hence width of the caustic network.

In all calculations we assumed a flat cosmological model, with $\Omega_{M}=0.27$, $\Omega_{\Lambda}=0.73$ and $H_{0}= 71\ \rm km\ s^{-1} Mpc^{-1}$.

\subsection{Parameters of source and lens}
\label{sec:param_src_lens}

Source is defined with the inner and outer radius. As we adopted the standard model for the disc we considered that the most of the radiation in the observed energy range, is coming from the disc part defined with the $R_{in}=10^{13}$m and $R_{out}=10^{15}$m, see \cite{Blackburne11}. For evaluating the proportionality coefficient $\alpha$, given by Eq. \ref{eqn:T0R0}, we took the temperature of
$T_0=2\cdot10^4$K, with the peak radiation around 150 nm, at the radius $R_0=3.15\cdot10^{13}$m. The coefficient $\beta$ usually has the value of $\beta=3/4$ and we kept that value constant throughout all simulations. The disc inclination can be changed, but in the most simulations we assumed a face-on disc orientation. The source plane dimension is the same in all our computation and equal to 40ERR.

The lens has been assumed to have a circular shape, containing $N_{s}$ stars of Solar mass ranging from 40 to 240. We also assumed that lens and source are placed at the cosmological distances, with $z_d=0.5$ and $z_s=2.0$ (standard lens). Those values are not constant in the case when we examine dependence on the source lens distance. We are confident, based on the discussion in the \S \ref{sec:mlmodel} that such carefully chosen lens reflect good enough condition for the gravitational bound systems. Any more massive and densely populated lens will act as one compact object with known influence on the distant source.

\section{Results and discussion}
\label{sec:results}

\begin{figure}
\centering
\includegraphics[width=8.5cm]{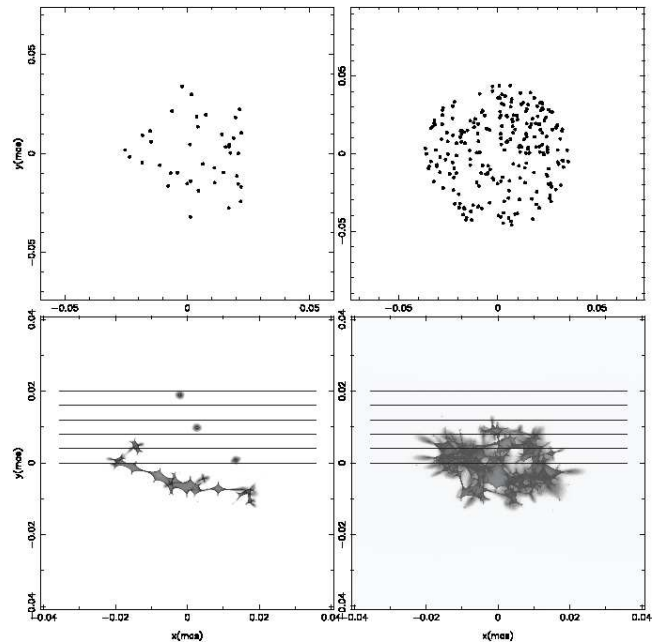}
\caption{Upper panels show a randomly star distribution in the lens plane, a) case of $n_{lens}=40$ (left panels) and b)
case for $n_{lens}=240$ (right panels). In the bottom panels we give the appropriate magnification map in the source plane for
specified $n_{lens}$, respectively. Horizontal parallel solid lines present the  paths of the  source for which we calculated magnification
light curves and variability of the photo-center positions. These plots are  given for a standard lens system with $z_d=0.5$ and $z_s=2.0$.}
\label{fig:stars_map}
\end{figure}

\begin{figure*}
\centering
\includegraphics[width=17.5cm]{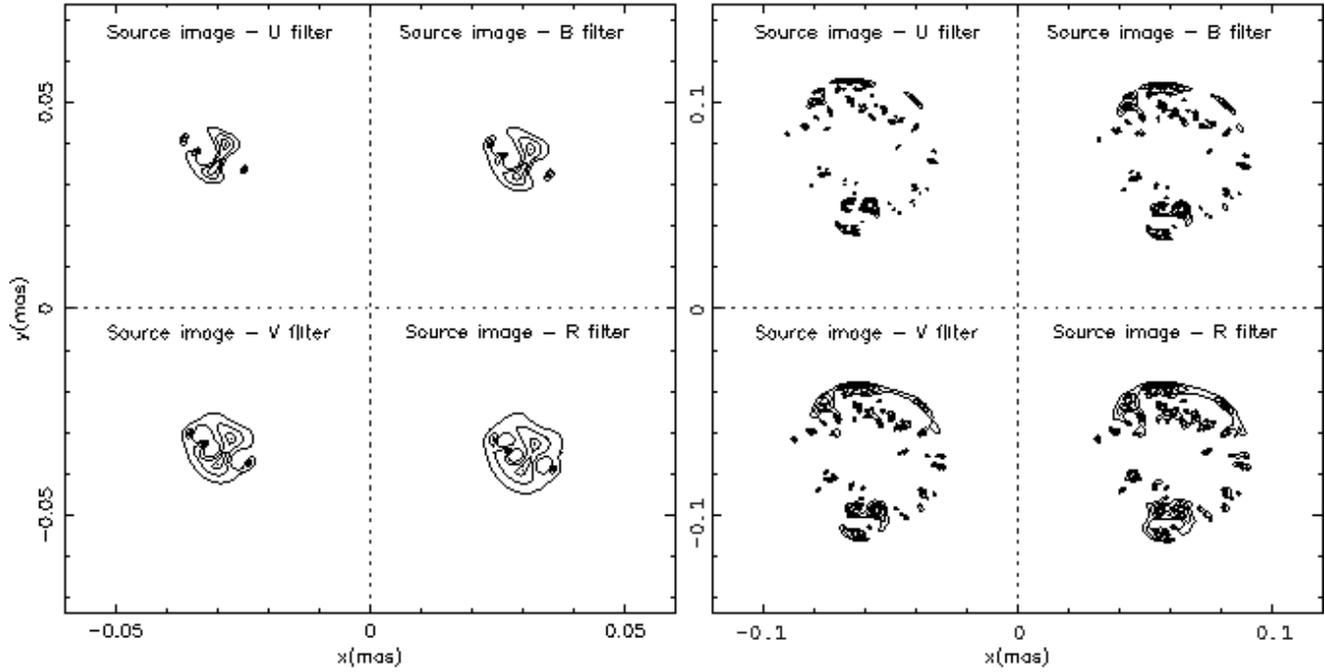}
\caption{Images of source in four different energy channels U[332-398nm], B[398-492nm], V[507-595nm] and R[589-727nm].
Lens for this case is presented in Figure \ref{fig:stars_map}. Left four panels present the image for the case of
$n_{lens}=40$, while four on the right side are for the $n_{lens}=240$ stars, with mentioned $z_d=0.5$ and $z_s=2.0$.}
\label{fig:map_gl}
\end{figure*}

As illustrations, we presented in Fig. \ref{fig:stars_map} the lens with random distribution of 40 and 240 stars (the distributions of stars located at $z_d$=0.5 are shown on upper panels, and proper magnification maps for a source at $z_s=2.0$ on bottom panels). As it can be seen from Fig. \ref{fig:stars_map} an increase of the star density brings that the lens is very similar to the point like one, therefore, we considered the maximal star number as 240 solar masses stars. The source images in U, B, V and R filters are shown in Fig. \ref{fig:map_gl}. As it can be seen the source images are different in dimension for different filters, and there is an expected magnification in size, that is increasing for the lower energies. Also, there is a noticeable change in the source structure due to the microlens effect that depends on the lensing system, as well as on the emission region of the source. In the case of $N_{s}=240$, which we considered as the massive lens, the images are composed from several bright similar structures in all filters, but one can see that the dimensions of bright structures are different. The images of almost point like source, in this case, are extensive and could be with dimensions of 0.1 mas (see Fig. \ref{fig:map_gl} right). It means that e.g. a non strongly lensed quasar, due to lensing by diffuse small mass structure, may show a complex shape with different bright spots in the center and arc-like structures on the edge.

\begin{figure*}
\centering
\includegraphics[width=17.5cm]{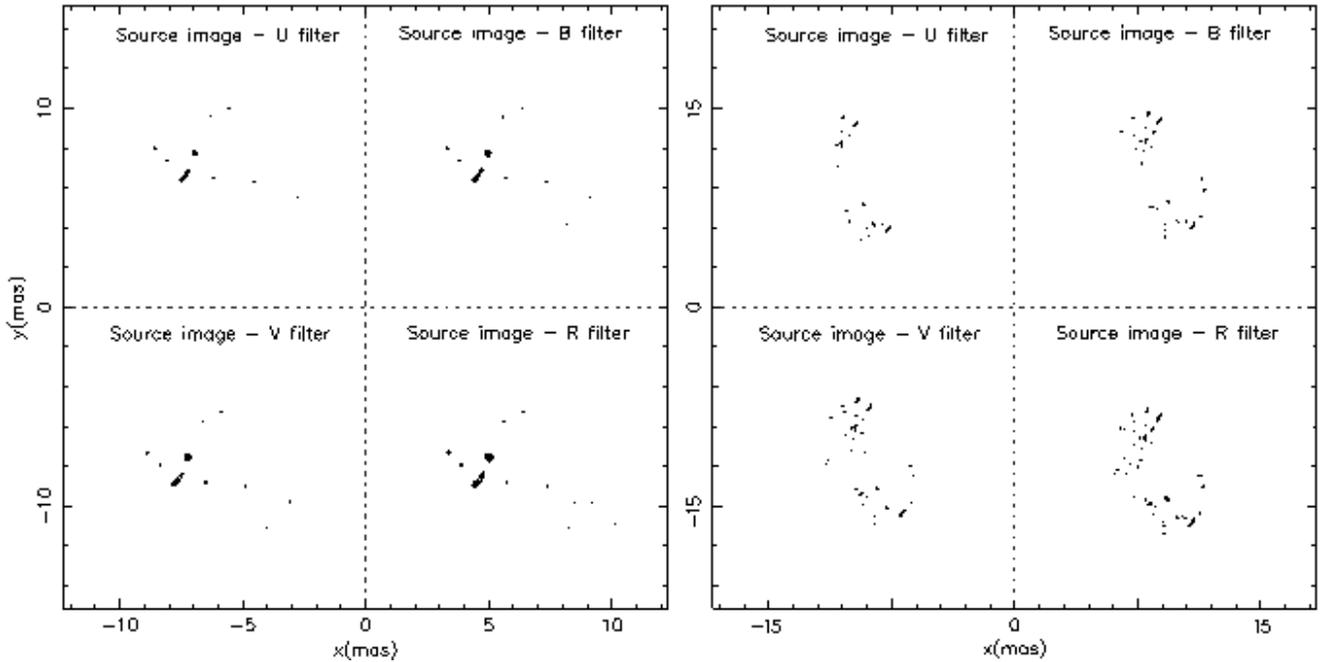}
\caption{The same as in Fig. \ref{fig:map_gl}, but for $z_d=0.01$ and $z_s=2.0$.}
\label{fig:map_gl1}
\end{figure*}

In some cases, one can expect that an image of the source is splitted into several components. Especially, in the case with low redshifted lens and high redshifted source. As an example in Fig. \ref{fig:map_gl1} we presented the images in different filters in the case where source is located at $z_s=2.0$ and lens at $z_d=0.01$. It is interesting that the lensed source can be seen as several images distributed across several milliarcseconds. This can affect a relatively strong jitter in the photocenter of a source, since random distribution of caustics contribute to the source amplification.

In the next chapter we present the influence of different parameters to the magnification and
photo-center offset in the case where the lens is  a stellar system with mass from 40 to 240 solar masses.

\subsection{Population of the stars in the lens}
\label{sec:spec_case_lens_param}

\begin{figure}
\centering
\includegraphics[width=8.5cm]{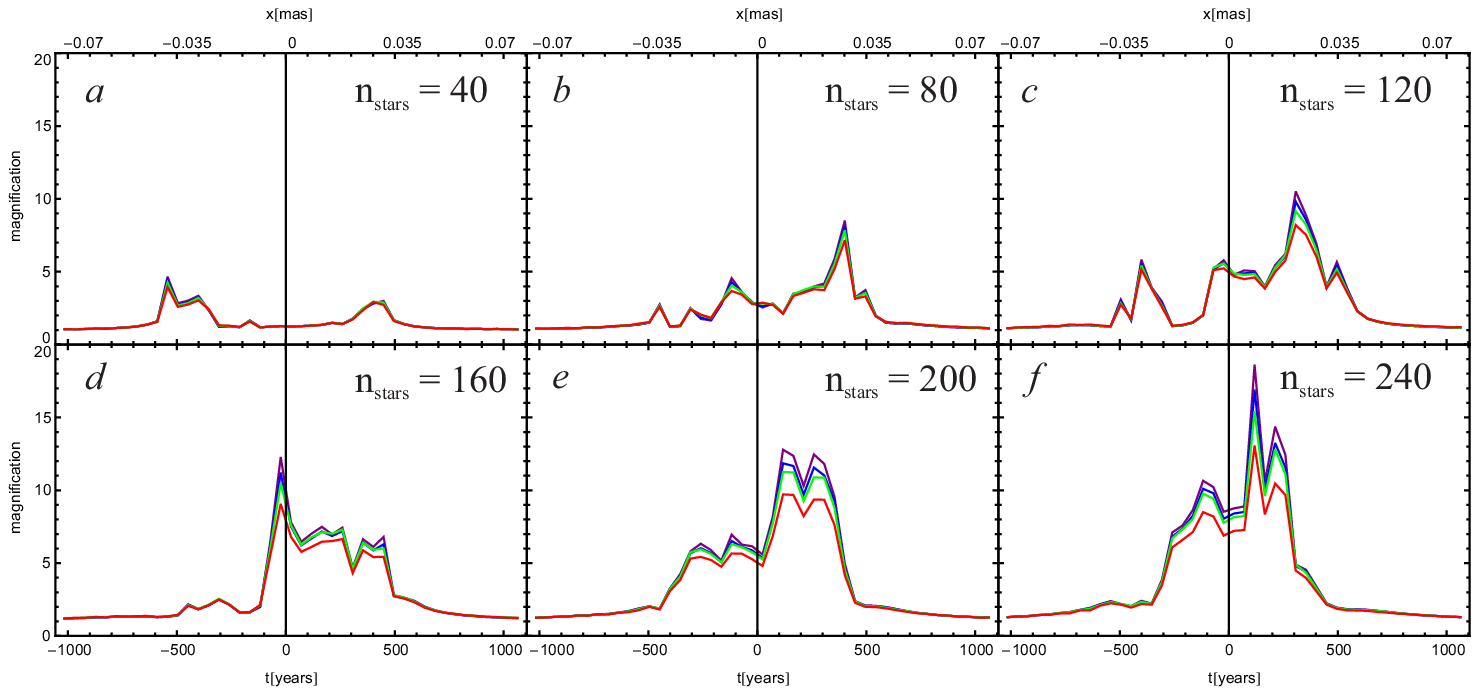}
\includegraphics[width=8.5cm]{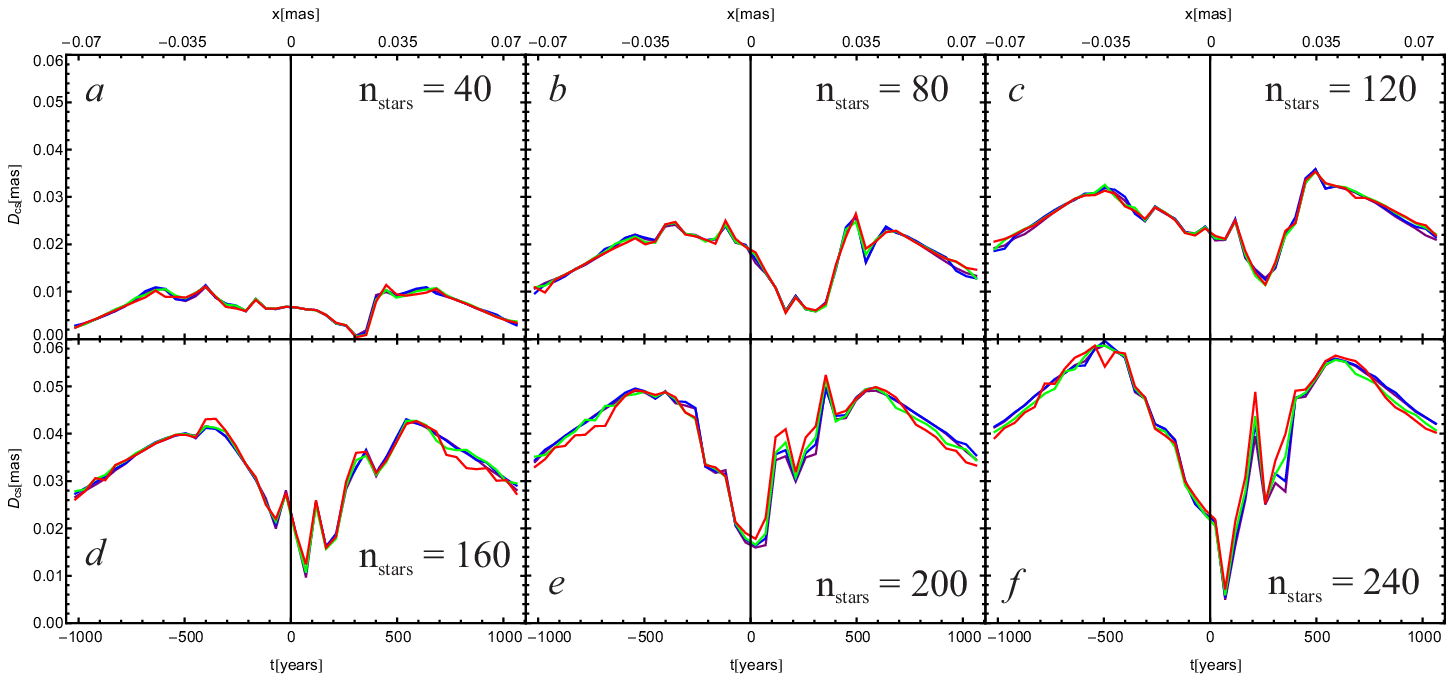}
\caption{Magnification (upper) and centroid shift (bottom) variation    for different star numbers in the lens. Different line colors represent observed energy channels, $U[332-398nm]$ - violet line, $B[398-492nm]$ - blue line, $V[507-595nm]$ - green line and $R[589-727nm]$ - red line.}
\label{fig:dfc_lum_snum}
\end{figure}

It is well known that a single lens microlensing of a source can cause a centroid motion, or wobbling of the photo-center of the source \citep[see e.g.][etc.]{Hog95,Walker95,Miyamoto95,Dominik00,Lee10}. Here, we considered an open cluster configuration of the lens, taking different number of stars, ranging from 40 to 240 Solar mass stars, on the same surface. We performed the simulations for a typical gravitational lens system, taking $z_d=0.5$ and $z_s=2.0$. The results of our simulations are presented in Fig. \ref{fig:dfc_lum_snum} and in Table \ref{tbl:snum}. Fig. \ref{fig:dfc_lum_snum} (upper panels) presents the variation of magnification for six different star numbers in the lens, ranging from 40 to 240. They are grouped and randomly distributed in a circular shape with diameter around 0.08 mas, as it is shown in Fig. \ref{fig:stars_map}. As it can be seen from Fig. \ref{fig:dfc_lum_snum} (upper panels), a significant difference between magnification in different filters is present for massive lens ($N_{s}\ge100$), and can be very prominent, i.e. the magnification in the U filter is significantly higher than in the R one, and as it can be seen in Table \ref{tbl:snum}, the magnification can be around 1.4 times higher in the U filter than in R one. This can also affect the flux anomaly between different images of a lensed quasar. On the other side, the light curves in different filters have similar shapes, and it is expected that with a higher lens mass (higher stellar density) the lens  acts as a point-like lens, i.e. the light curves have a Gaussian-like shape (see Fig. \ref{fig:dfc_lum_snum}, upper -- panel right-down). In the case of low density star clusters we can distinguish separate peaks in the magnification curve induced by the absence of caustics over the source path.

The photocenter variability for considered cases is given in Fig. \ref{fig:dfc_lum_snum} (bottom panels) and maximal photocenter offset in Table \ref{tbl:snum}.  As it is expected, the photocenter variability is higher for massive lens. Fig. \ref{fig:dfc_lum_snum} (bottom panels) also shows that in almost all cases a decrease of the photocenter offset with approaching the center of the diffuse mass lens is seen, and the drop is higher for the massive lens. As it can be seen in Fig.\ref{fig:dfc_lum_snum} (bottom panels) and  Table \ref{tbl:snum} in difference with the magnification, there is no significant photocenter offset difference among different filters with exception for the last two cases of 200 and 240 stars. We conclude that for cluster with higher star population this mutual band deviation will not increase significantly, regarding the single lens behavior of the lens. The maximal photo-center offset is small, around 0.06 mas for the case of a high density cluster (see Table \ref{tbl:snum}).

\begin{table*}
\centering
\noindent
\small{
\caption{Data for the maximal centroid shift in $[mas]$ and magnification per energy channel for
different number of stars in the lens, ranging from 40 to 240.
First column gives the number of micro lenses $n_{lens}$, next four are centroid shift in U, B, V and R energy
channels, and next four are magnification variability in same channels. Calculations are made for the standard lens and source distances.}
\begin{tabular}{|>{\centering\arraybackslash}m{1cm}|>{\centering\arraybackslash}m{1cm}|>{\centering\arraybackslash}m{1cm}|>{\centering\arraybackslash}m{1cm}|>{\centering\arraybackslash}m{1cm}|
>{\centering\arraybackslash}m{1cm}|>{\centering\arraybackslash}m{1cm}|>{\centering\arraybackslash}m{1cm}|>{\centering\arraybackslash}m{1cm}|>{\centering\arraybackslash}m{1cm}|}
\hline
\multirow{3}{*}\emph{$n_{lens}$} & \emph{$\kappa$} & \multicolumn{4}{c|}{\textbf{$d_{fc}[mas]$}} & \multicolumn{4}{c|}{\textbf{$magn.[-]$}} \\
- & - & U & B & V & R & U & B & V & R \\
\hline
40 & 0.0940 & 0.011 & 0.011 & 0.011 & 0.011 & 4.6 & 4.4 & 4.2 & 3.9 \\
\hline
80 & 0.1620 & 0.026 & 0.025 & 0.024 & 0.026 & 8.5 & 8.1 & 7.8 & 7.1 \\
\hline
120 & 0.2182 & 0.035 & 0.035 & 0.035 & 0.035 & 10.5 & 9.8 & 9.1 & 8.2 \\
\hline
160 & 0.2650 & 0.042 & 0.043 & 0.042 & 0.043 & 12.2 & 11.2 & 10.4 & 9.0 \\
\hline
200 & 0.3070 & 0.050 & 0.049 & 0.051 & 0.052 & 12.7 & 11.8 & 11.2 & 9.7 \\
\hline
240 & 0.3440 & 0.058 & 0.059 & 0.058 & 0.058 & 18.6 & 16.9 & 15.4 & 13.0 \\
\hline
\end{tabular}
\label{tbl:snum}
}
\end{table*}

\subsubsection{Different lens and source redshifts}

It is interesting to explore a situation where star clusters, containing between several tens to 240 stars, are located at low redshift and act as lenses to different redshifted quasars. Therefore, we performed numerical simulations, assuming that the lens is a low redshifted star cluster ($z_d=0.01$). The linear dimension of the lens is held constant, consequently when moved closer to the observer it's angle diameter is increased. The lens contains 80 Solar mass stars concentrated in a circular shape with diameter 2.7 mas. The distance of a source is changed from $z_s=0.05$ to $z_s=2.0$.

\begin{figure}
\centering
\includegraphics[width=8.5cm]{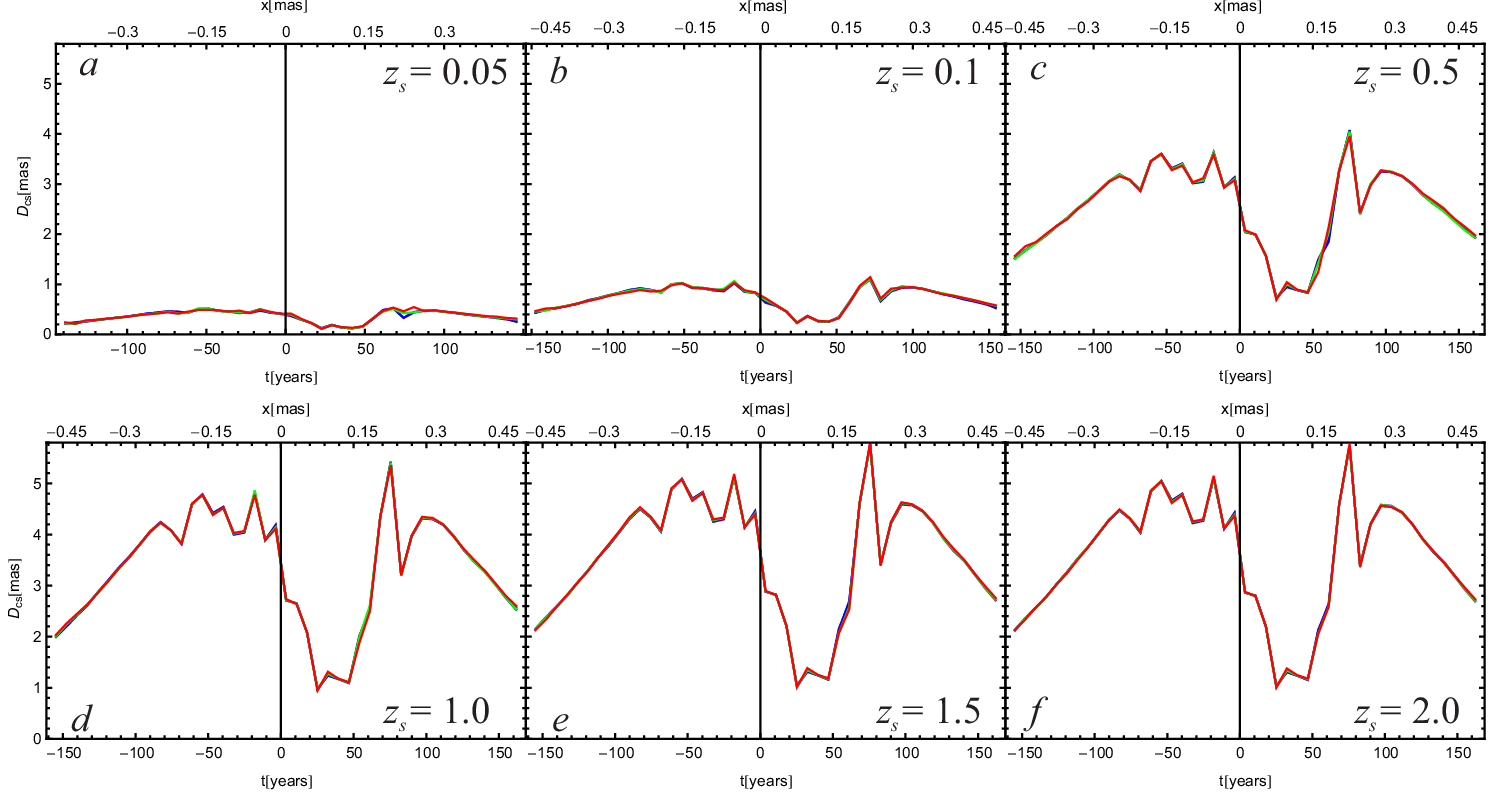}
\includegraphics[width=8.5cm]{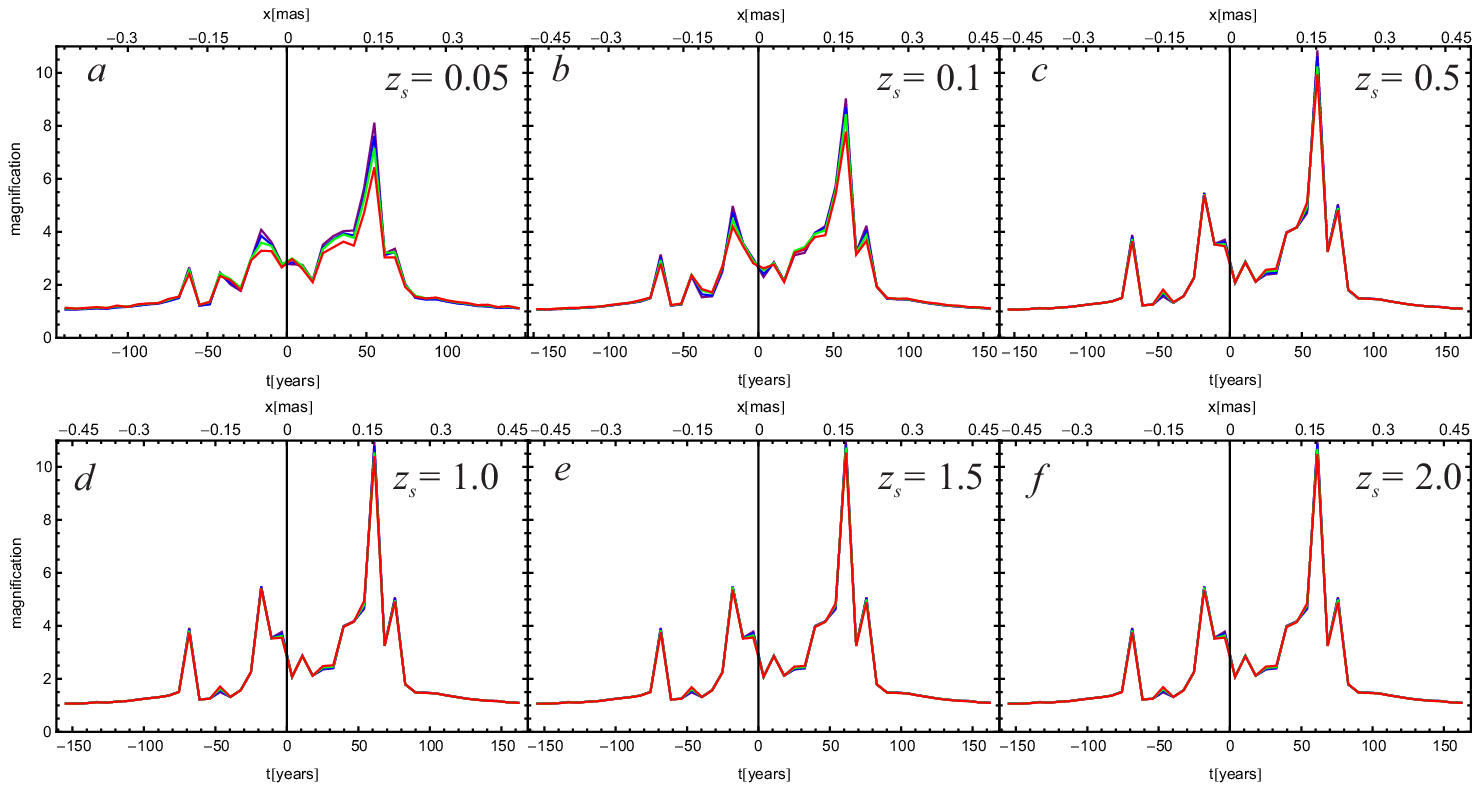}
\caption{Same as in the Figure \ref{fig:dfc_lum_snum}, but for different values of source redshift $z_s$ and the
lens contains 80 Solar mass stars.}
\label{fig:dfc_lum_zs_all}
\end{figure}

Results of simulations are presented in Fig. \ref{fig:dfc_lum_zs_all}, and maximal photo-center offset and magnification in Table \ref{tbl:zs}. As it can be seen from Fig. \ref{fig:dfc_lum_zs_all} and   Table \ref{tbl:zs}, the amplification is almost constant after $z_s=0.5$ source redshift and photocenter offset also stays almost the same, taking values around 3-5 mas. Therefore, one can notice that a low redshifted star cluster acting as a lens will produce similar amplification and photocenter offset on all high redshifted quasars. In all cases a flux anomaly is presented and as it is expected, due to dimension of the UV emitting region, the highest amplification is seen in the U filter.

On the other hand, the time scales of the lens passing the lens across the source are smaller, consequently a high variability in the flux, can be detected in a period of several years (see Fig. \ref{fig:dfc_lum_zs_all}). This indicates that high redshifted sources projected very close to the center of a low redshifted galaxy can be affected by gravitational microlensing by small mass clusters, and therefore, they are not suitable candidates for astrometry.

\begin{table*}
\centering
\noindent
\small{
\caption{Same as in the Table \ref{tbl:snum}, but for different values of source distance defined with $z_s$, ranging
from 0.05 to 2.0. The results are given for the case when the  lens contains 80 Solar mass stars ($\kappa = 0.162$)
located at close distance of $z_d=0.01$.}
\begin{tabular}{|>{\centering\arraybackslash}m{1cm}|>{\centering\arraybackslash}m{1cm}|>{\centering\arraybackslash}m{1cm}|>{\centering\arraybackslash}m{1cm}|
>{\centering\arraybackslash}m{1cm}|>{\centering\arraybackslash}m{1cm}|>{\centering\arraybackslash}m{1cm}|>{\centering\arraybackslash}m{1cm}|>{\centering\arraybackslash}m{1cm}|}
\hline
\multirow{3}{*}{\emph{$z_{s}$}} & \multicolumn{4}{c|}{\textbf{$d_{fc}[mas]$}} & \multicolumn{4}{c|}{\textbf{$magn.$}} \\
& U & B & V & R & U & B & V & R \\
\hline
0.05 & 0.52 & 0.51 & 0.52 & 0.54 & 8.11 & 7.61 & 7.15 & 6.43 \\
\hline
0.1 & 1.11 & 1.10 & 1.10 & 1.14 & 9.03 & 8.69 & 8.44 & 7.77 \\
\hline
0.5 & 4.07 & 4.07 & 4.04 & 3.95 & 10.84 & 10.61 & 10.24 & 9.94 \\
\hline
1.0 & 5.42 & 5.41 & 5.40 & 5.35 & 10.93 & 10.78 & 10.55 & 10.41 \\
\hline
1.5 & 5.78 & 5.76 & 5.75 & 5.78 & 10.95 & 10.84 & 10.72 & 10.53 \\
\hline
2.0 & 5.74 & 5.74 & 5.74 & 5.77 & 10.96 & 10.87 & 10.67 & 10.48 \\
\hline
\end{tabular}
\label{tbl:zs}
}
\end{table*}

\subsection{Influence of impact parameter of the microlensing event}
\label{sec:spec_case_impc_param}

All above simulations have been performed for a source crossing the center of a stellar cluster (lens), but there is possibility that the projected distance between the lens and source are different. Therefore we simulated cases of different impact parameters \emph{b}, for a small mass (40 stars) and more massive (240 stars)  lens, taking a standard lens system ($z_d=0.5$ and $z_s=2.0$).

The parameter \emph{b} is given in relative units in comparison with the source plane half height and has been changed from 0 (when the lens and source are centered) to 0.5 (lens located at the half height of the lens plane). We simulated transition of the lens taking paths as presented in Fig. \ref{fig:stars_map} (horizontal lines, the line crossing the center corresponds to \emph{b}=0 and the outer one to \emph{b}=0.5).

\begin{figure}
\centering
\includegraphics[width=8.5cm]{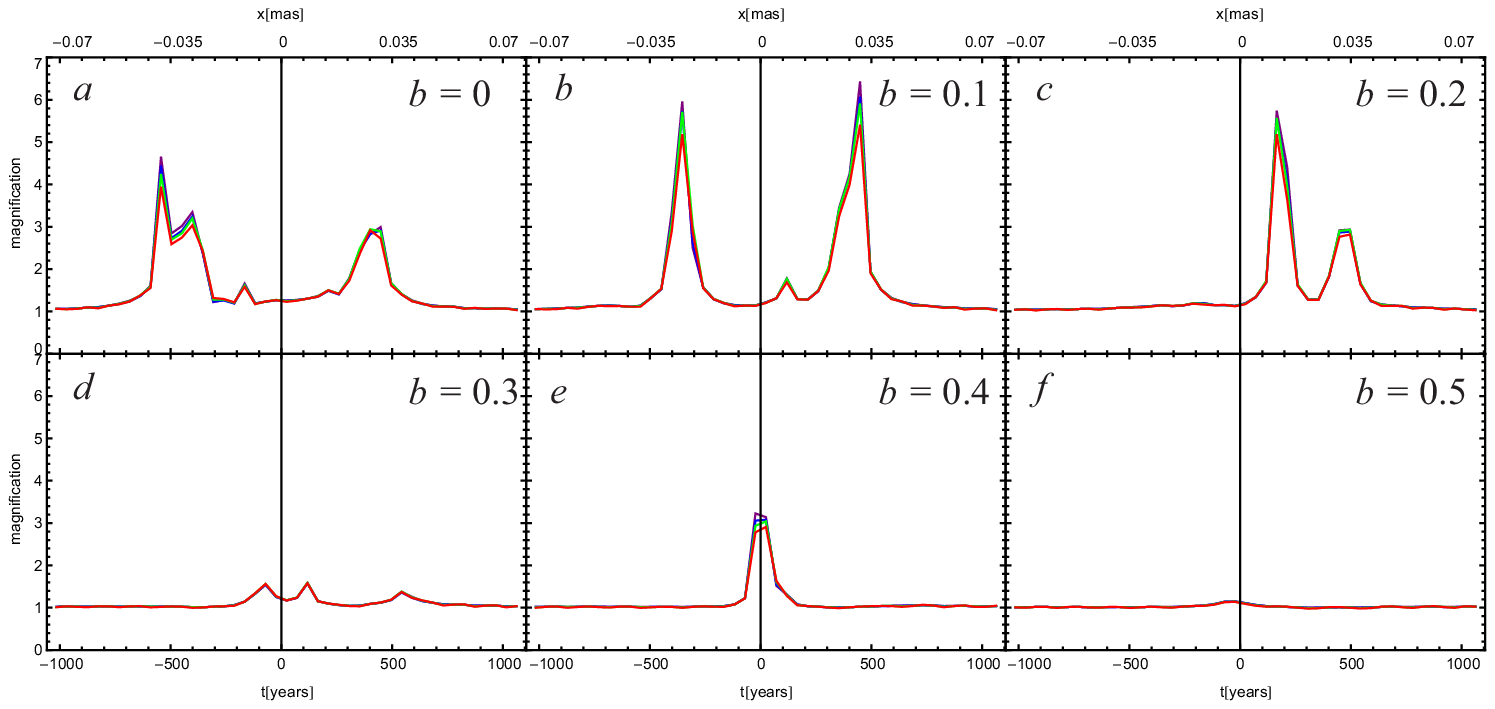}
\includegraphics[width=8.5cm]{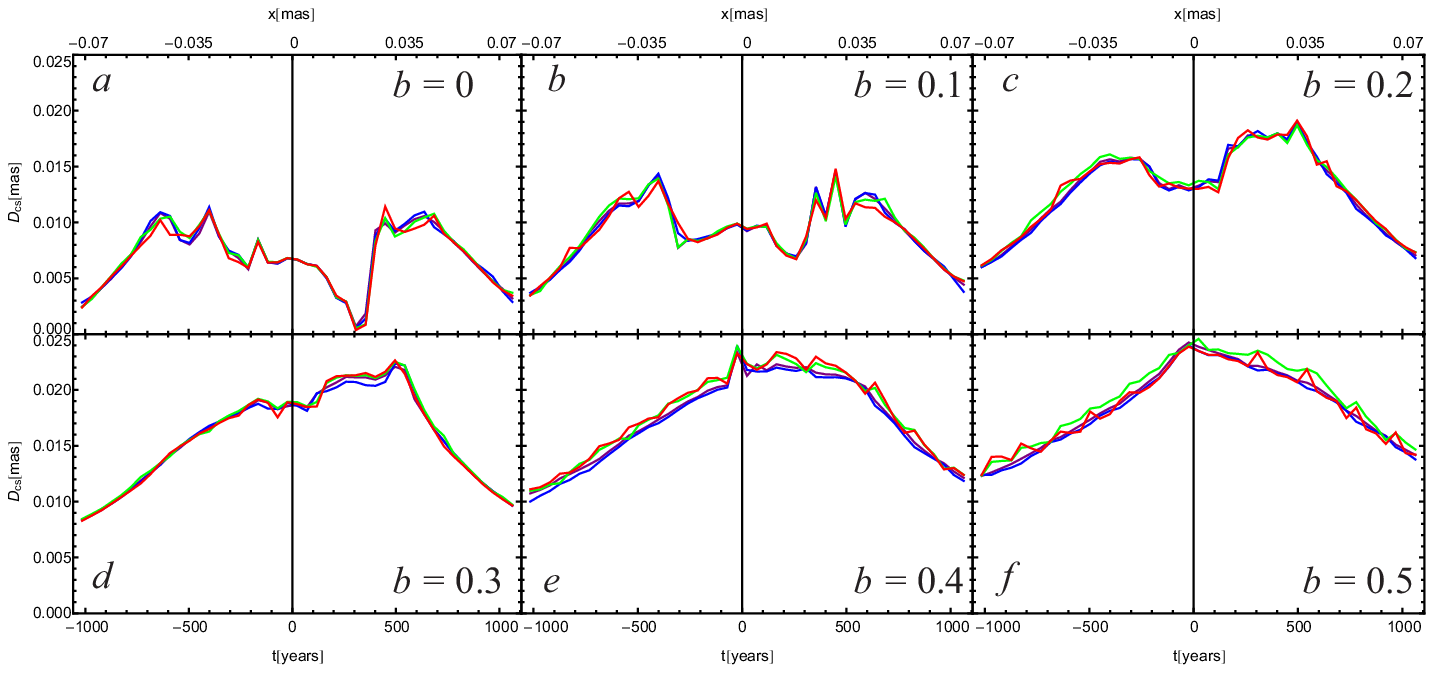}
\caption{Magnification (upper panel) and photocenter (bottom panel) variability of a source during the overcrossing event. Lens consist of 40 Solar mass stars distributed uniformly in the circular area with the diameter of 0.04mas. In panels from a) to f) impact parameter b changes from 0 towards maximum value equal to the half of the halfheight of source map. Trajectories are presented as in the Figure \ref{fig:stars_map}. Distances $D_d$ and $D_s$ are defined with it's respectful redshifts with values for the standard lens ($z_d=0.5$ and $z_s=2.0$).}
\label{fig:dfc_lum_impc_40}
\end{figure}

\begin{figure}
\centering
\includegraphics[width=8.5cm]{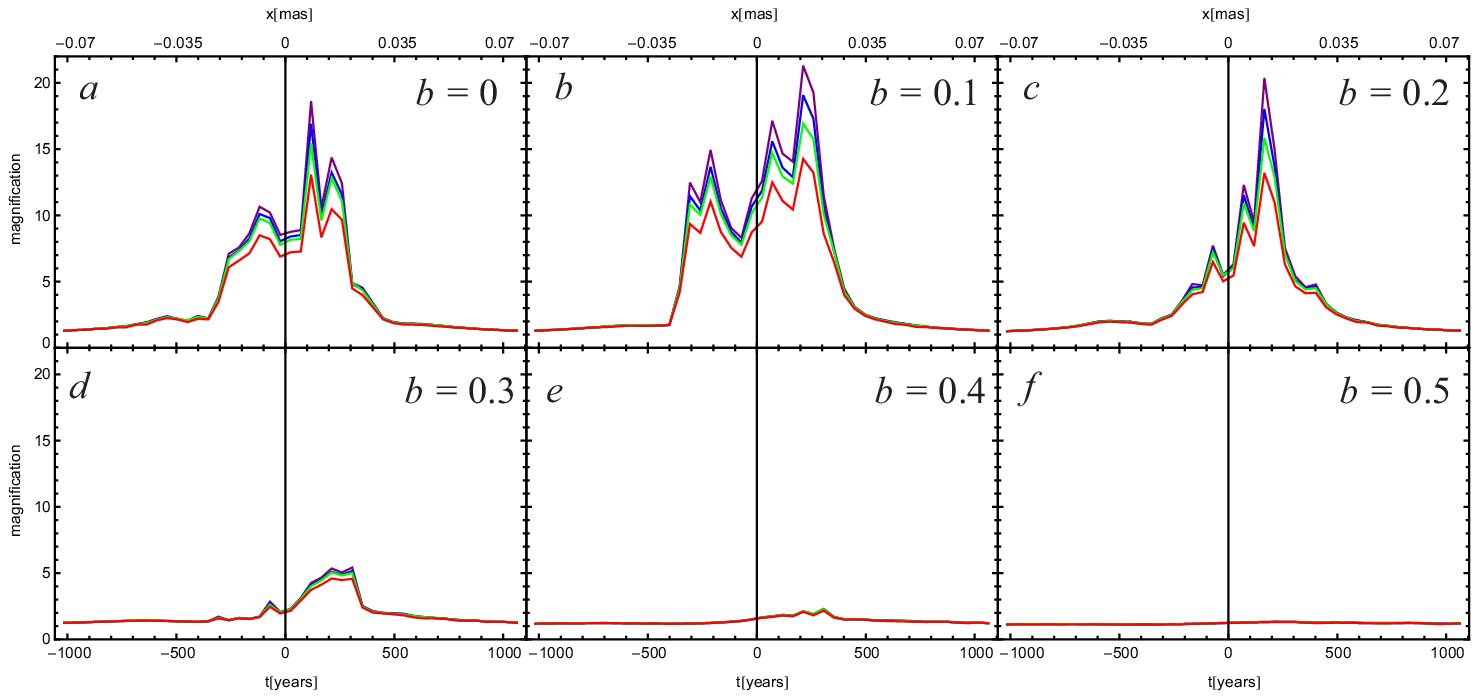}
\includegraphics[width=8.5cm]{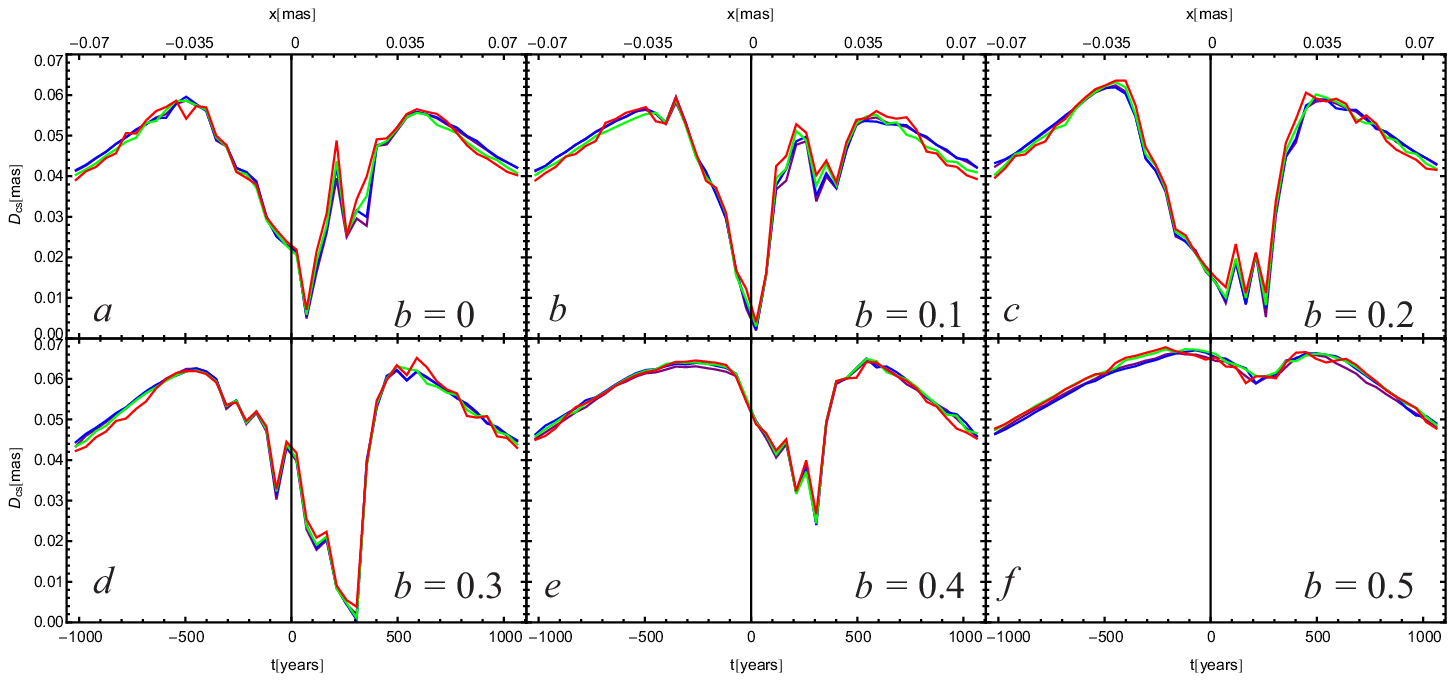}
\caption{Same as in the Fig. \ref{fig:dfc_lum_impc_40}, but for cluster of 240 Solar mass stars.}
\label{fig:dfc_lum_impc_240}
\end{figure}

For this simulations we selected two extreme cases with 40 and 240 stars in the lens in order to have a clear picture of what happens during the variation of this parameter. We showed them in the form of Table \ref{tbl:impc} and Figs. \ref{fig:dfc_lum_impc_40} and \ref{fig:dfc_lum_impc_240}. For example, in  Fig. \ref{fig:dfc_lum_impc_40} (upper panel) we see that for the case of lens with 40 stars, there are more or less similar diagrams with casual peaks of the magnification having similar height. Those are caused by a rare caustic distribution in the magnification map for a small number of stars. One can conclude that one peak in  lightcurves presents a variation caused by a single caustic in the map (caused by a group of stars), with relative variation as much as 5 to 7 times. On the contrary, in the second case (Fig. \ref{fig:dfc_lum_impc_240}), with a more massive lens (240 stars) we have the proportionally higher relative magnification of the order of 20 times, caused by an action of a bulk of overlapping caustics which gradually decreases with an increase of parameter $b$. For the last two cases ( e) and f) in  Fig. \ref{fig:dfc_lum_impc_240}) the magnification is neglected. Such behavior suggests that the  lens acts as a compact object with the mass equal to the sum of  masses of all stars in the cluster. In both cases relative deviations per observed energy channels is at maximum for the moment of the highest magnification with the maximum in the U channel and  decreased for another channels. These observed properties are more clearly given in Table \ref{tbl:impc}, column $magn$.

For the case of the centroid shift variable $D_{cs}$ presented in Figs. \ref{fig:dfc_lum_impc_40} and \ref{fig:dfc_lum_impc_240} (bottom panels) for both cases of 40 and 240 stars we can notice a lot of similarities with main difference in the amount of shift which is 2 to 3 times higher, for latter. For the both cases we have two peaks which are evolving with increasing of the parameter $b$.

The most interesting is that in the case of low mass cluster, where one can expect almost
same magnification and photocenter variation in all passbands form b=0 to b=0.5. Such behavior is caused since
randomly distributed groups of stars can be homogenously distributed within the cluster.

\begin{table*}
\centering
\noindent
\small{
\caption{Same as in the Table \ref{tbl:snum}, but for different values of impact parameters of the source and lens,
ranging from 0 to 0.5 of map half height. Paths of source over the magnification map are presented with full lines
in  Figure \ref{fig:stars_map}. We give the results for two cases when lens contain 40 and 240 stars, ($\kappa = 0.094$ and $\kappa = 0.344$ respectively).}
\begin{tabular}{|>{\centering\arraybackslash}m{0.5cm}|>{\centering\arraybackslash}m{0.5cm}|>{\centering\arraybackslash}m{1cm}|>{\centering\arraybackslash}m{1cm}|>{\centering\arraybackslash}m{1cm}|
>{\centering\arraybackslash}m{1cm}|>{\centering\arraybackslash}m{1cm}|>{\centering\arraybackslash}m{1cm}|>{\centering\arraybackslash}m{1cm}|>{\centering\arraybackslash}m{1cm}|}
\hline
\multirow{3}{*}{\emph{b}} & \multicolumn{1}{c|}{\textbf{$n_{stars}$}} & \multicolumn{4}{c|}{\textbf{$d_{fc} [mas]$}} & \multicolumn{4}{c|}{\textbf{$magn.$}} \\
& - & U & B & V & R & U & B & V & R \\
\hline
0 & 40 & 0.010 & 0.011 & 0.011 & 0.011 & 4.6 & 4.4 & 4.2 & 3.9 \\
\hline
0 & 240 & 0.058 & 0.059 & 0.058 & 0.058 & 18.6 & 16.9 & 15.4 & 13.0 \\
\hline
0.1 & 40 & 0.014 & 0.014 & 0.014 & 0.014 & 6.4 & 6.1 & 5.9 & 5.4 \\
\hline
0.1 & 240 & 0.058 & 0.059 & 0.058 & 0.059 & 21.3 & 19.1 & 16.9 & 14.2 \\
\hline
0.2 & 40 & 0.018 & 0.018 & 0.018 & 0.019 & 5.7 & 5.5 & 5.5 & 5.2 \\
\hline
0.2 & 240 & 0.062 & 0.061 & 0.063 & 0.063 & 20.3 & 18.0 & 15.8 & 13.2 \\
\hline
0.3 & 40 & 0.022 & 0.022 & 0.022 & 0.022 & 1.6 & 1.5 & 1.6 & 1.5 \\
\hline
0.3 & 240 & 0.062 & 0.062 & 0.063 & 0.065 & 5.4 & 5.1 & 5.0 & 4.5 \\
\hline
0.4 & 40 & 0.023 & 0.023 & 0.023 & 0.023 & 3.2 & 3.1 & 3.0 & 2.9 \\
\hline
0.4 & 240 & 0.064 & 0.065 & 0.064 & 0.064 & 2.3 & 2.3 & 2.3 & 2.1 \\
\hline
0.5 & 40 & 0.024 & 0.023 & 0.024 & 0.023 & 1.1 & 1.1 & 1.1 & 1.1 \\
\hline
0.5 & 240 & 0.066 & 0.067 & 0.067 & 0.067 & 1.3 & 1.3 & 1.3 & 1.3 \\
\hline
\end{tabular}
\label{tbl:impc}
}
\end{table*}

\section{Conclusions}
\label{sec:concl}

In this paper we studied the microlensing effect caused by a small mass (diffuse) stellar cluster (up to 240 solar masses) defined in the \S \ref{sec:mlmodel} on the compact QSO source as it is defined in \S \ref{sec:srcmodel}. The case of such lensing may be present in the case of  a gravitational lens system, where one of images can be affected by the  lensing of a low mass cluster. Moreover, this can be present in the case of quasars which are not strongly lensed by the central mass of a galaxy, but they are projected close to the center of a low redshifted galaxy. In the most cases, we considered a typical lens system ($z_d=0.5$ and $z_s=2.0$) for a low mass cluster having 40 or 80 Solar mass stars in a circular shaped surface with diameter of around 0.08 mas. We explored situations with different distances of a source, taking that the lensing stellar cluster is low red shifted ($z_d=0.01$). We investigated  variations of magnification and photo-center in different (U,B,V,R) spectral bands, covering the spectral range that will be observed with the Gaia space mission. From our investigations, we are able to point out the  following conclusions:

i) A small mass stellar cluster (from 40 to 240 solar mass stars) can significantly contribute to the flux and photo-center variability, not only in the case of lensed quasars, but also in the case of quasars which are projected very close to the center of a low redshifted galaxy. The amplification is different in different spectral filters, ranging from 5 to 20 times; the UV spectrum is more amplified than the IR one. On the other side, the photo-center variability can reach several milliarcseconds, that can be detected with future missions.

ii) The lensing of such cluster can significantly contribute to the flux anomaly observed in some lensed quasars. In this anomaly one can expect that the flux in the  U filter will be always more amplified than in the R filter. This shape of amplification can be used to clarify the source of the anomaly. On the other hand, the photocenter offset is not significantly different in different filters. In the case of $N_L\le 100$ lensing, one can expect that the image of a lensed quasar or non-lensed quasar has a structure showing brighter spots around the center and arc-like structures on the edge of the projected Einstein ring of the cluster. Extremely, in the case of low redshifted lens ($z_d\le0.01$) and high redshifted source ($z_d\ge0.5$), there may be several  features (or point like images) separated by several milliarcseconds.

iii) In a difference with the point like lens, the diffuse stellar cluster microlensing can produce several prominent peaks in light curve during the lensing event, especially in a low mass  cluster (up to 100 solar mass stars distributed in the considered surface of 0.08 mas). It depends on surface stellar distribution across the projected surface of the cluster, and in the low mass stellar cluster, the caustics caused by grouping of stars can significantly affect the amplification (different in different spectral filters), as well as cause an offset the photocenter.

Additionally, we should emphasize a special case with a low redshifted small mass cluster ($z_d\le0.01$) in combination with high redshifted quasars ($z_d\ge0.5$). The effect of lensing in this case is very important, the light of quasars may be amplified in a relatively  short period (several years), and photocenter variability can be detected in a relatively short time. Thus, quasars projected very close to the low redshifted galaxies might not be good candidates for the reference frame objects. In this case one can expect maximal offset of the photocenter of several milliarcseconds, that is comparable with the offset of the photocenter caused by the changes in the inner quasar structure \citep[see][]{pop12}.

\section*{Acknowledgments}

This work is a part of the project (176001) "Astrophysical Spectroscopy of Extragalactic Objects," supported by the Ministry of Science  and Technological Development of Serbia. We would like to thank the anonymous referee for very useful comments and suggestions.


\begin{thebibliography}{99}




\bibitem[\protect\citeauthoryear{Belokurov \& Evans}{2002}]{Belokurov02} Belokurov, V. A \& Evans, N. W., \ 2002, MNRAS, 331, 649

\bibitem[\protect\citeauthoryear{Blackburne et al.}{2006}]{bl05} Blackburne, J. A., Pooley, D., Rappaport, S. \ 2006, ApJ, 640, 569

\bibitem[\protect\citeauthoryear{Blackburne et al.}{2011}]{Blackburne11}  Blackburne, J. A.,  Pooley, D.,  Rappaport,  S. \&
Schechter, P. L. \ 2011, ApJ, 729, 34

\bibitem[\protect\citeauthoryear{Boden et al.}{1998}]{Boden98} Boden, A. F., Shao, M. \& van Buren, D., \ 1998, ApJ, 502, 538

\bibitem[\protect\citeauthoryear{Chen et al.}{2007}]{ch07} Chen, J., Rozo, E., Dalal, N., Taylor, J. E. \ 2007, ApJ, 659, 52

\bibitem[\protect\citeauthoryear{Congdon \& Keeton}{2005}]{co05} Congdon, A. B., Keeton, C. R. \ 2005, MNRAS, 364, 1459

\bibitem[\protect\citeauthoryear{Dalal \& Lane}{2003}]{Dalal03} Dalal, N. \& Lane, B. F., \ 2003, ApJ, 589, 199

\bibitem[\protect\citeauthoryear{Delplancke et al.}{2001}]{Delplancke01} Delplancke, F., Gorski, K. M., \& Richichi, A., \ 2001, A\&A., 375, 701

\bibitem[\protect\citeauthoryear{Dominik \& Sahu}{2000}]{Dominik00} Dominik, M. \& Sahu, K., \ 2000, ApJ, 534, 213

\bibitem[\protect\citeauthoryear{Done et al.}{2011}]{Done11} Done C., Davis, S.W., Jin, C, Blaes, O. \& Ward, M., \ 2011, MNRAS, 420, 1848D

\bibitem[\protect\citeauthoryear{Erickcek \& Law}{2011}]{Erick11} Erickcek, A. L. \& Law, N. M. \ 2011, ApJ, 729, 49

\bibitem[\protect\citeauthoryear{Goldberg \& Wozniak}{1998}]{Goldberg98} Goldberg, D. M., \& Wozniak, P. R., \ 1998, Acta Astron., 48, 19

\bibitem[\protect\citeauthoryear{Han \& Kim}{1999}]{Han99} Han, C., \& Kim, T. W., \ 1999, MNRAS, 305, 795

\bibitem[\protect\citeauthoryear{Hog et al.}{1995}]{Hog95} Hog, E., Novikov, I. D. \& Polnarev, A. G., \ 1995, A\&A, 294, 287

\bibitem[\protect\citeauthoryear{Honma \& Kurayama}{2002}]{Homna02} Honma, M., \& Kurayama, T., \ 2002, ApJ, 568, 717

\bibitem[\protect\citeauthoryear{Hosokawa et al.}{1997}]{Hosokawa97} Hosokawa, M., Ohnishi, K. \& Fukushima, T., \ 1997, AJ, 114, 1508

\bibitem[\protect\citeauthoryear{Inoue \& Chiba}{2005}]{in05} Inoue, K. T., Chiba, M. \ 2005, ApJ, 634, 77

\bibitem[\protect\citeauthoryear{Jovanovi\'c  et al.}{2008}]{jov08} Jovanovi\'c, P., Zakharov, A. F., Popovi\'c, L. \v C., Petrovi\'c, T. \ 2008, MNRAS, 386, 397


\bibitem[\protect\citeauthoryear{Kayser et al.}{1986}]{Kayser86} Kayser, R., Refsdal, S. \& Stabell, R. \ 1986, A\&A, 166, 36



\bibitem[\protect\citeauthoryear{Keeton \& Moustakas}{2009}]{ke09} Keeton, C. R., Moustakas, L. A. \ 2009, ApJ, 699, 1720

\bibitem[\protect\citeauthoryear{Kratzer et al.}{2011}]{kr11} Kratzer, R. M., Richards, G. T., Goldberg, D. M. et. al. \ 2011, ApJ, 728L, 18

\bibitem[\protect\citeauthoryear{Krolik}{1998}]{Krolik98} Krolik, J. H. \ 1998, Active Galactic Nuclei: From the Central Black Hole to the
Galactic Environment (Princeton University Press)


\bibitem[\protect\citeauthoryear{Lee et al.}{2010}]{Lee10} Lee, C. H., Seitz, S., Riffeser, A. \& Bender, R., \ 2010, MNRAS, 407, 1597

\bibitem[\protect\citeauthoryear{Mao \& Witt}{1998}]{Mao98} Mao, S. \& Witt, H. J., \ 1998, MNRAS, 300, 1041

\bibitem[\protect\citeauthoryear{Meusinger et al.}{2010}]{me10} Meusinger, H., Henze, M., Birkle, K., Pietsch, W., et al. \ 2010, A\&A, 512A, 1

\bibitem[\protect\citeauthoryear{Mignard et al.}{2002}]{Mignard02} Mignard F., \ 2002, in Bienayme O., Turon C., eds, Proc. GAIA: A European
Space Project, Vol. 2, held on 2001 May 14-18, Les Houches, France. EDP Sciences, EAS Publications Series, p. 327

\bibitem[\protect\citeauthoryear{Miralda-Escude}{1996}]{Miralda96} Miralda-Escude, J., \ 1996, ApJ, 470, L113

\bibitem[\protect\citeauthoryear{Miyamoto \& Yoshii}{1995}]{Miyamoto95} Miyamoto, M. \& Yoshii Y., \ 1995, AJ, 110, 1427

\bibitem[\protect\citeauthoryear{Novikov \& Thorne}{1973}]{Novikov73} Novikov, I.D. \& Thorne, K.S., \ 1973, blho.conf, 343

\bibitem[\protect\citeauthoryear{Oguri}{2005}]{og05} Oguri, M. \ 2005, MNRAS, 361L, 38

\bibitem[\protect\citeauthoryear{Paczynski}{1998}]{Paczinski98} Paczynski, B., \ 1998, ApJ, 494, L23

\bibitem[\protect\citeauthoryear{Petters et al}{2001}]{Pett01} Petters A. O., Levine H. \& Wambsganss J., \ 2001, Singular Theory and Gravitational
Lensing, Birkhšauser, Boston


\bibitem[\protect\citeauthoryear{Poindexter et al.}{2008}]{Poindexter08} Poindexter, S., Morgan, N. \& Kochanek, C. S., \ 2008, ApJ, 673, 34

\bibitem[\protect\citeauthoryear{Pooley et al.}{2006}]{po06} Pooley, D., Blackburne, J. A., Rappaport, S., Schechter, P. L., Fong,\ 2006, ApJ, 648, 67

\bibitem[\protect\citeauthoryear{Pooley et al.}{2012}]{po12} Pooley, D., Rappaport, S., Blackburne, J. A., Schechter, P. L., Wambsganss, J. \ 2012, ApJ, 744, 111

\bibitem[\protect\citeauthoryear{Popovi\'c \& Chartas}{2005}]{pop05} Popovi\'c, L. \v C., Chartas, G. \ 2005, MNRAS, 357, 135

\bibitem[\protect\citeauthoryear{Popovi\'c et al.}{2012}]{pop12} Popovi\'c, L. \v C.,; Jovanovi\'c, P., Stalevski, M., Anton, S., Andrei, A. H., Kova\v cevi\'c, J., Baes, M. \ 2012, A\&A, 538A, 107

\bibitem[\protect\citeauthoryear{Pringle \& Rees}{1972}]{Pringle72} Pringle, J.E. \& Rees, M.J., \ 1972, A\&A, 21, 1

\bibitem[\protect\citeauthoryear{Proft et al.}{2011}]{pr11} Proft, S., Demleitner, M., Wambsganss, J. 2011, A\&A, 536A, 50

\bibitem[\protect\citeauthoryear{Safizadeh et al.}{1999}]{Safizadeh99} Safizadeh, N., Dalal, N., \& Griest, K., \ 1999, ApJ, 522, 512

\bibitem[\protect\citeauthoryear{Sazhin et al.}{2011}]{Sazhin11} Sazhin, M. V., Sazhina, O. S., Pshirkov, M. S. 2011, Ast. Rep, 55, 954

\bibitem[\protect\citeauthoryear{Sazhin et al.}{1998}]{Sazhin98} Sazhin, M. V., Zharov, V. E., Volynkin, A. V., \& Kalinina, T. A., 1998, \ MNRAS, 300, 287


\

\bibitem[\protect\citeauthoryear{Schmidt \&  Wambsganss}{2010}]{sc10} Schmidt, R. W.\& Wambsganss, J. \ 2010, GReGr, 42, 2127

\bibitem[\protect\citeauthoryear{Schneider et al}{1992}]{sch92} Schneider, P., Ehlers, J., \& Falco, E.E. \ 1992, Gravitational Lenses,
Springer-Verlag Berlin -- Heidelberg -- New York


\bibitem[\protect\citeauthoryear{Schneider \& Weiss}{1986}]{Schneider86} Schneider, P. \& Weiss, A. \ 1986, A\&A, 164, 237

\bibitem[\protect\citeauthoryear{Schneider \& Weiss}{1987}]{Schneider87} Schneider, P. \& Weiss, A. \ 1987, A\&A, 171, 49

\bibitem[\protect\citeauthoryear{Shakura \& Sunyaev}{1973}]{Shakura73} Shakura, N.I. \& Sunyaev, R.A., \ 1973, A\&A, 24, 337

\bibitem[\protect\citeauthoryear{Sluse et al.}{2012}]{sl12} Sluse, D., Hutsem\'ekers, D., Courbin, F., Meylan, G., Wambsganss, J.
2012, A\&A, 544A, 62

\bibitem[\protect\citeauthoryear{Stalevski et al.}{2012}]{st12} Stalevski, M., Jovanovi\'c, P., Popovi\'c, L. \v C., Baes, M.
2012, MNRAS.425.1576

\bibitem[\protect\citeauthoryear{Treyer \& Wambsganss}{2004}]{Treyer04} Treyer, M. \& Wambsganss, J. \ 2004, A\&A, 416, 19

\bibitem[\protect\citeauthoryear{Treu}{2010}]{tr10} Treu, T. \ 2010, ARA\&A, 48, 87

\bibitem[\protect\citeauthoryear{Walker}{1995}]{Walker95} Walker, M. A., \ 1995, ApJ, 453, 37

\bibitem[\protect\citeauthoryear{Williams \& Saha}{1995}]{Williams95} Williams, L. L. R. \& Saha, P., \ 1995, AJ, 110, 1471

\bibitem[\protect\citeauthoryear{Zackrisson \& Riehm}{2007}]{za07} Zackrisson, E., Riehm, T. \ 2007, A\&A, 475, 453

\bibitem[\protect\citeauthoryear{Zackrisson \& Riehm}{2010}]{za10} Zackrisson, E., Riehm, T.\ 2010, AdAst, 2010, ID 478910

\bibitem[\protect\citeauthoryear{Zaharov et al}{2004}]{zah04} Zakharov, A. F., Popovi\'c, L. \v C., Jovanovi\'c, P.\ 2004, A\&A, 420, 881

\bibitem[\protect\citeauthoryear{Zakharov}{1997}]{Zah97}Zakharov A. F., \ 1997, Gravitational Lenses and Microlensing, Janus-K, Moscow

\bibitem[\protect\citeauthoryear{Yano}{2012}]{ya12} Yano, T. 2012, ApJ, 757, 189


\end{thebibliography}
\end{document}